\documentclass[10pt]{iopart}	
\usepackage{amssymb,bm}
\usepackage{calrsfs}
\usepackage{cite}
\usepackage {graphicx}

\usepackage{hyperref}
\hypersetup{
    colorlinks=true,        
    linkcolor=blue,          
    citecolor=blue,        
    filecolor=magenta,      
    urlcolor=blue           
}
\usepackage{color} 				
\definecolor{dgray}{rgb}{0.6,0.6,0.6}
\definecolor{dmag}{rgb}{0.6,0.0,0.6}
\usepackage[normalem]{ulem} 	
\newcommand{\ao}[1]{{#1}}

\definecolor{pink}{rgb}{1,0,0.9}

\newcommand{\eqref}[1]{{(\ref{#1})}}
\newcommand{\pdg}{{\phantom{\dagger}}}

\usepackage{simpler-wick}
\begin{document}

\title[Thermodynamics of a weakly interacting ...]{Thermodynamics of a weakly interacting Bose gas above the transition temperature}

\author{M S Bulakhov$^{1,2}$, A S Peletminskii$^{1,2}$, Yu~V~Slyusarenko$^{1,2}$, and A G Sotnikov$^{1,2}$}
\address{$^{1}$ Akhiezer Institute for Theoretical Physics, NSC KIPT, 61108 Kharkiv, Ukraine}
\address{$^{2}$ V.N. Karazin Kharkiv National University, 61022 Kharkiv, Ukraine}
\ead{bulakh@kipt.kharkov.ua}


\date{\today}

\begin{abstract}
We study thermodynamic properties of weakly interacting Bose gases above the transition temperature of Bose-Einstein condensation in the framework of a thermodynamic perturbation theory.
Cases of local and non-local interactions between particles are analyzed both analytically and numerically.
We obtain and compare the temperature dependencies for the chemical potential, entropy, pressure, and specific heat to those of noninteracting gases.
The results set reliable benchmarks for thermodynamic characteristics and their asymptotic behavior in dilute atomic and molecular Bose gases above the transition temperature.
\end{abstract}


\maketitle
\ioptwocol	

\section{Introduction}

For more than two decades, the study of the physical properties of quantum Bose gases has been the subject of intensive research, both theoretical and experimental \ao{\cite{Pethick,Stringari, Haugset1998, Andersen2004}}. This is due to experimental realization of Bose-Einstein condensation (BEC) in dilute atomic gases of alkali metals \cite{Anderson,Davis,Bradley}, which became possible after the elaborated techniques on laser cooling and trapping of neutral atoms. The experiments have confirmed a number of predictions made by theory of a weakly interacting Bose gas with condensate developed by Bogoliubov \cite{Bogoliubov,ZB}. 
It became almost the first theory, where it was necessary to essentially abandon the methods of the standard thermodynamic perturbative approach when describing the interaction effects. This is because of the appearance of divergent terms in the series of thermodynamic perturbation theory at small values momentum. At the same time, the Bogoliubov's theory itself contains nonanalyticity in the interaction \cite{Tolmachev}, which does not allow to perform a consistent expansion in interaction as in the standard thermodynamic perturbative approach. Its another weak point is that it is valid at extremely low temperature, close to absolute zero.

While a majority of theoretical studies were (reasonably) focused on the temperature regime considerably below the transition to \ao{the BEC state~\cite{Pethick,Stringari, Haugset1998, Andersen2004}}, the proximity of the transition itself and the region above the critical point in weakly interacting Bose gases is much less studied.
In particular, existing theoretical approaches provide no definite answer on basic questions, such as how the critical temperature depends on the amplitude of the interaction potential or how the thermodynamic quantities evolve 
\ao{above the transition temperature}
in comparison with the ideal Bose gases.

In this paper, we attempt to describe thermodynamics of a Bose gas above the transition temperature to determine the structure of interaction corrections to the main thermodynamic quantities. The theoretical description is constructed in the framework of the thermodynamic perturbation theory (TPT) introduced first by Peierls \cite{Peierls1933, Matsubara1955, Landau5eng}.
While in context of weakly interacting quantum gases this theoretical approach was suggested long time ago \cite{Abrikosov,Akh-Pel}, to the best of our knowledge, it has never been applied to obtain specific characteristics of a weakly interacting Bose gas or to quantitatively analyze its thermodynamic quantities.

\section{Formalism}
Let us consider an equilibrium system described by the Gibbs statistical operator
\begin{equation}
    w = \exp{ \left[ \beta 
     \left( \Omega - \mathcal{H} - \mu N \right) 
     \right] },
\end{equation}
where $\beta$ is the reciprocal temperature in energy units ($\beta=1/T$), $\Omega$ is the thermodynamic potential, $\mu$ is the chemical potential and $N$ is the particle number operator.
We assume that the system Hamiltonian~$\mathcal{H}$ can be split into two parts, such that
\begin{equation}
    \mathcal{H}
     = \mathcal{H}_0
     + \mathcal{V},
\end{equation}
where $\mathcal{H}_0$ is the free-particle term and $\mathcal{V}$ describes the interaction between particles.
In order to proceed, we introduce the following operator:
\begin{equation}\label{eq:K}
    K \left( \beta \right) 
     = e^{ \beta \left( \mathcal{H}_0 
     - \mu N \right) } 
     e^{ -\beta \left( \mathcal{H} - \mu N \right) }.
\end{equation}
Then, from the normalization condition for $w$ (${\rm Tr}\,w=1$), one can determine the thermodynamic potential of a system of weakly interacting particles,
\begin{equation}
    \Omega 
     = \Omega_0
     - \ln{ \left< K \left( \beta \right) \right>_0 }/\beta ,
\end{equation}
where 
$ \left< A \right>_0 \equiv \Tr w_0 A$, 
$ w_0 = \exp{ \left\{ \beta \left( \Omega_0 - \mathcal{H}_0 - \mu N \right) \right\} } $,
and the zero subscripts correspond to the system of noninteracting particles (an ideal gas).

It is important to note now that the operator $K\left(\beta\right)$ can be expanded into series in $\beta\mathcal{V}$ \cite{Akh-Pel},
\begin{equation}
    K  
     = \sum\limits_{n=0}^{\infty} \frac{ \left( -1 \right) ^n }{ n! } 
     \int\limits_0^{\beta} d\lambda_1 \dots \int\limits_0^{\beta} d\lambda_n 
     T 
     \left[ 
      \mathcal{V}\left(\lambda_1\right) \dots \mathcal{V}\left(\lambda_n\right) 
     \right],
\end{equation}
where
\begin{equation}
  \mathcal{V} \left( \lambda_i \right) 
  =  e^{\lambda_i \left( \mathcal{H}_0 - \mu N \right)} 
   \mathcal{V} 
   e^{ -\lambda_i \left( \mathcal{H}_0-\mu N \right)}
\end{equation}
and $T$ is the ordering operator for the variable~$\lambda$.

The obtained formula can be further simplified by summing over the uncoupled terms, i.e., those which contain a group of multipliers 
$\mathcal{V}\left(\lambda_{i_1}\right) \dots \mathcal{V}\left(\lambda_{i_k}\right)$ 
coupled only with each other, in the calculation of expectation values according to the Wick's theorem.
\ao{In other words, the uncoupled terms of the second or higher expansion order are those that do not contain any operators entering $\mathcal{V}(\lambda_{a})$ entangled with entries of $\mathcal{V}(\lambda_{b})$ by contractions (for pairwise interaction, see below).
}
This yields \cite{Abrikosov,Akh-Pel},
\begin{equation}\label{eq:omc}
    \Omega = 
    \Omega_0 - 
    \left( \left< K \left( \beta \right) \right>_0^c - 1 \right)/\beta
\end{equation}
with
\begin{eqnarray}\label{eq:Kc}
    \left< K \left( \beta \right) \right>_0^c 
    &=& \sum\limits_{n=0}^{\infty} \frac{ \left( -1 \right) ^n }{ n! } 
    \int\limits_0^{\beta} d\lambda_1 
    \dots \int\limits_0^{\beta} d\lambda_n 
    \nonumber
    \\
    &&\times\left< T \left[ \mathcal{V}\left(\lambda_1\right) \dots \mathcal{V}\left(\lambda_n\right) \right] \right>_0^c,
\end{eqnarray}
where the index $c$ denotes that the coupled terms are accounted only.

Below, we focus on the properties of a weakly interacting Bose gas governed
by the Hamiltonian that can be expressed in the momentum space as
\begin{equation}
    \mathcal{H}_0
    =
    \sum\limits_{\bf p}
    \varepsilon_{\bf p}^\pdg
    a_{\bf p}^\dagger
    a_{\bf p}^\pdg,
\end{equation}
\begin{equation}
    \mathcal{V}
    =
    \frac1{2V}
    \sum_{ {\bf p}_1 \dots {\bf p}_4 }
    \nu_{{\bf p}_1{\bf p}_3}^\pdg
    a_{{\bf p}_1}^\dagger
    a_{{\bf p}_2}^\dagger
    a_{{\bf p}_3}^\pdg
    a_{{\bf p}_4}^\pdg
    \delta_{ {\bf p}_1 + {\bf p}_2 , {\bf p}_3 + {\bf p}_4}.
\end{equation}
where $a_{{\bf p}}^\dagger$ ($a_{{\bf p}}$) is the bosonic creation (annihilation) operator of a particle with the momentum~${\bf p}$, $V$ is the system volume, $\nu_{{\bf p}_1{\bf p}_3} \equiv \nu({\bf p}_1-{\bf p}_3)$ is the Fourier transform of the pairwise interaction,
and $\varepsilon_{\bf p}$ is the free-particle energy. 
\ao{In particular, by specifying $\mathcal{V}$ in terms of creation and annihilation operators, the uncoupled terms are those of the type \cite{Abrikosov}}
\[
\langle\mathcal{V}_1\cdot\mathcal{V}_2\rangle^{uc}\propto
\wick{
    \c1{a}_{{\bf p}_1}^\dagger
    \c2{a}_{{\bf p}_2}^\dagger
    \c2{a}_{{\bf p}_3}^\pdg
    \c1{a}_{{\bf p}_4}^\pdg
    }
    \cdot
    \wick{
    \c1{a}_{{\bf p}_5}^\dagger
    \c2{a}_{{\bf p}_6}^\dagger
    \c1{a}_{{\bf p}_7}^\pdg
    \c2{a}_{{\bf p}_8}^\pdg
    }+\ldots, 
\]
\ao{
while the coupled are those of the type
}
\[
\langle\mathcal{V}_1\cdot\mathcal{V}_2\rangle^{c}\propto
\wick{
    \c1{a}_{{\bf p}_1}^\dagger
    \c2{a}_{{\bf p}_2}^\dagger
    \c3{a}_{{\bf p}_3}^\pdg
    \c1{a}_{{\bf p}_4}^\pdg
    \cdot
    \c1{a}_{{\bf p}_5}^\dagger
    \c3{a}_{{\bf p}_6}^\dagger
    \c1{a}_{{\bf p}_7}^\pdg
    \c2{a}_{{\bf p}_8}^\pdg
    }+\ldots~. 
\]

For simplicity and definiteness, we consider a three-dimensional gas consisting of the nonrelativistic spinless particles with the mass~$m$ and $\varepsilon_{\bf p}=p^2/2m$.
In the noninteracting limit, the gas is described by the standard Bose-Einstein distribution function
\begin{equation}\label{eq:BE-distr}
  f_{p} = \left\{ \exp[\beta \left( \varepsilon_p - \mu \right)] -1 \right\} ^{-1}.
\end{equation}

We also introduce the corresponding transition temperature to the BEC state in the ideal Bose gas,
\begin{equation}\label{eq:BE-T0}
  T_0 = \frac{1}{2m}
  \left[
    \frac{4\pi^2\hbar^3n}{\Gamma(3/2)\zeta(3/2)}
  \right]^{2/3},
\end{equation}
where $\hbar$ is the Planck's constant, $n=N/V$ is the density of a gas, $\Gamma(x)$ and $\zeta(x)$ are the Gamma and Riemann zeta functions, respectively.
\ao{Note that throughout this study the density $n$ is kept fixed, thus}
$T_0$ serves below as a scaling unit in all temperature and energy-related dependencies. \ao{In contrast, the effect of finite number of particles and the trap curvature in the weakly-intaracting Bose gas above the transition temperature can be found, e.g., in Refs.~\cite{Biswas2009,Goswami2013}.}

It is also insightful to introduce an additional scaling quantity in the coordinate space, namely the thermal de Broglie wavelength of particles at $T=T_0$,
\begin{equation}\label{eq:BE-L0}
  \Lambda_0 = \sqrt{
  \frac{2\pi\hbar^2}{mT_0}
  },
\end{equation}
which is of the order of the average distance between the closest two particles of a gas, $\Lambda_0\sim n^{-1/3}$.

\section{Main thermodynamic quantities}

\subsection{Chemical potential}

To determine the TPT corrections to the chemical potential and main thermodynamic quantities of a weakly interacting Bose gas we address Eqs.~(\ref{eq:omc}), (\ref{eq:Kc}). It is convenient to start from the expansion of the thermodynamic potential into series up to the quadratic corrections in interaction,
\begin{eqnarray}
  \Omega(\mu) = \Omega_0(\mu) + \Omega_1(\mu) + \Omega_2(\mu) + \dots,
  \label{eq:Om}
  \\
  \Omega_0(\mu) = - T\sum\limits_{{\bf p}} \ln{ \left( 1 + f_p \right) },
  \\
  \Omega_1(\mu) 
  = \left< \mathcal{V} \right> _0^c 
  = \frac{ 1 }{ 2V } \sum\limits_{{\bf p}{\bf p}^\prime} \left(\nu_0 + \nu_{ {\bf p} {\bf p}^\prime }\right) f_{p} f_{p^{\prime}},
  \\
  \Omega_2(\mu)
   = \Omega_2^\prime(\mu) + \Omega_2^{\prime\prime}(\mu),
   \label{eq:Om2}
\end{eqnarray}
where \ao{$\nu_0\equiv\nu({\bf p}-{\bf p}'=0)$, while}
\begin{eqnarray*}
  \Omega_2^\prime(\mu) 
   = - \frac{ \beta }{ 2V^2 } 
   \sum\limits_{{\bf p}_1{\bf p}_2{\bf p}_3}   
   \left[ 
    \nu_0^2 
    + 2 \nu_0 \nu_{{\bf p}_2{\bf p}_3} + \nu_{{\bf p}_1{\bf p}_3} 
    \nu_{{\bf p}_2{\bf p}_3} 
   \right]
   \\
   \qquad\qquad\qquad\times
   f_{p_1} f_{p_2} f_{p_3} \left( f_{p_3} + 1 \right),
  \\
  \Omega_2^{\prime\prime}(\mu) 
  = - \frac{ \beta }{ 4V^2 } \sum\limits_{{\bf p}_1\dots{\bf p}_4}
    \left[
        \nu_{{\bf p}_1{\bf p}_3}^2 + \nu_{{\bf p}_1{\bf p}_3} \nu_{{\bf p}_1{\bf p}_4}
    \right]
   \\
    \qquad\qquad\times
    f_{p_1} f_{p_2} \left( f_{p_3} + 1 \right) \left( f_{p_4} + 1 \right) 
    \delta_{ {\bf p}_1 + {\bf p}_2 , {\bf p}_3 + {\bf p}_4 }.
\end{eqnarray*}
Here $\delta$ is the Kronecker symbol.

Let us determine the total number of particles~$N$ from the Maxwell relation,
\begin{equation}\label{eq:Mrel}
  N = - \left( 
   \frac{ \partial \Omega }{ \partial \mu } 
   \right)_{T,V} .
\end{equation}
According to the expansion \eqref{eq:Om}, one obtains
\begin{eqnarray}
  N(\mu) = N_0(\mu) + N_1(\mu) + N_2(\mu) + \dots,
  \label{eq:N}
  \\
  N_0(\mu) = \sum\limits_{\bf p} f_p,
  \\
  N_1(\mu) = - \frac{ \beta }{ V } \sum\limits_{{\bf p}{\bf p}^\prime} 
  \left[ \nu_0 + \nu_{ {\bf p} {\bf p}^\prime } \right] 
  f_{p^{\prime}} f_p \left( f_p + 1 \right),
  \label{eq:N1}
  \\
  N_2(\mu) = N_2^\prime(\mu) + N_2^{\prime\prime}(\mu),
  \label{eq:N2}
\end{eqnarray}
where
\begin{eqnarray*}
  N_2^\prime(\mu)
    =
    \frac{ \beta^2 }{ 2V^2 }
    \sum\limits_{{\bf p}_1{\bf p}_2{\bf p}_3}
        \left[
            \nu_0^2 + 2 \nu_0 \nu_{{\bf p}_2{\bf p}_3} + \nu_{{\bf p}_1{\bf p}_3} \nu_{{\bf p}_2{\bf p}_3}
        \right]
    \\
    \qquad\times
    f_{p_1} f_{p_2} f_{p_3} \left( f_{p_3} + 1 \right)
    \left(
        2 f_{p_3} + f_{p_1} + f_{p_2} + 3
    \right),
  \\
  N_2^{\prime\prime}(\mu)
    =
    \frac{ \beta^2 }{ 4V^2 }
    \sum\limits_{ {\bf p}_1 \dots {\bf p}_4 }
    \left[
        \nu_{{\bf p}_1{\bf p}_3}^2
        +
        \nu_{{\bf p}_1{\bf p}_3}
        \nu_{{\bf p}_1{\bf p}_4}
    \right]
    \\
    \qquad\times
    f_{p_1} f_{p_2}
    \left( f_{p_3} + 1 \right)
    \left( f_{p_4} + 1 \right)
    \\
    \qquad\times
    \delta_{{\bf p}_1 + {\bf p}_2 , {\bf p}_3 + {\bf p}_4 }
    \left(
        f_{p_1} + f_{p_2} + f_{p_3} + f_{p_4} + 2
    \right).
\end{eqnarray*}

Note that the thermodynamic quantities of interest are the functions of the chemical potential~$\mu$. Therefore, for a consistent perturbative approach, it is necessary to expand the chemical potential up to the second order in interaction, 
\begin{eqnarray*}
	\mu(T,V,N)
	= \mu_0 + \mu_1 + \mu_2 + \dots,
\\
	\mu_1
	\equiv
	\sum\limits_{{\bf p}}
	\left.
	\left(
		\frac{
			\partial
			\mu
		}{
			\partial
			\nu_{\bf p}
		}
	\right)
	_{T,V,N}
	\right|
	_{0
	}
	\nu_{{\bf p}}
,
\\
	\mu_2
	\equiv
	\frac12
	\sum\limits_{{\bf pp}^\prime}
	\left.
	\left(
		\frac{ \partial^2 \mu }{ \partial \nu_{{\bf p}} \partial \nu_{{\bf p}^\prime} }
	\right)_{T,V,N}
	\right|
	_{0}
	\nu_{{\bf p}}
	\nu_{{\bf p}^\prime},
\end{eqnarray*} 
where the subscript 0 denotes the substitution $\mu=\mu_0 $ and \ao{$\nu_{\bf p}\equiv\nu({\bf p})=0$} after taking the derivative.

The chemical potential $\mu_0 $ is determined from the normalization condition for the total number of particles~$N$ (see, e.g., Refs.~\cite{Pathria2011,chempot2017ltp} for details).
Then, the introduced higher-order corrections $\mu_1 $ and $\mu_2$ are obtained from Eq.~\eqref{eq:N} by imposing the condition $\partial N/\partial \nu_{\bf p}=0$ and using the relations
$N_1=\sum_{\bf p}(\partial N/\partial \nu_{\bf p}) \nu_{\bf p}$ and $N_2=\frac{1}{2}\sum_{\bf p}(\partial^2 N/\partial \nu_{\bf p}\partial\nu_{\bf p'}) \nu_{\bf p}\nu_{\bf p'}$.
After a few mathematical transformations, we obtain
\begin{eqnarray}
	\mu_1
	=
	- N_1 ( \mu_0 ) \left( \frac{ \partial N_0 ( \mu_0 ) }{ \partial \mu_0 } \right)^{-1} _{T,V}
	\label{eq:mu}
,\\
	\mu_2
	=
	- \left( \frac{ \partial N_0 ( \mu_0 ) }{ \partial \mu_0 } \right)^{-1} _{T,V} 
	\left[ 
	 \left( \frac{ \partial^2 N_0 ( \mu_0 ) }{ \partial \mu_0^2 } \right) _{T,V} \frac{\mu_1^2}{2}
	\right.\nonumber
  \\
  \left.\qquad
	+
	\left( \frac{ \partial N_1 ( \mu_0 ) }{ \partial \mu_0 } \right)_{T,V} \mu_1
	+
	N_2 ( \mu_0 )
	\right]
	.
	\label{eq:mu2}
\end{eqnarray}

In case of local interaction, i.e., the momentum-independent Fourier transform $\nu_{\bf pp'}=\nu_0$, the corrections~\eqref{eq:mu} and \eqref{eq:mu2} take the form
\begin{equation}\label{eq:muloc}
	\mu_{1}^{(loc)}
	=
	2 n \nu_0
,\quad
	\mu_{2}^{(loc)}
	=
	-
	N_{2}^{\prime\prime}
	(\mu_0)
	\left(
	\frac{ \partial N_0 ( \mu_0 ) }{ \partial \mu_0 }
	\right)^{-1}_{T,V}.
\end{equation}

Before analysing the TPT corrections to the chemical potential, let us specify the form of the employed nonlocal interaction potential.
In accordance with the previous studies \cite{Bulakhov2018}, it is convenient to employ the model of semitransparent spheres with the Fourier transform
\begin{equation}
	\nu_{\bf p}
	=
	3g
	\frac{
		j_1(pr_0/\hbar)
	}{
		pr_0/\hbar
	},
\end{equation}
where $ g = 4\pi \hbar^2 a_s / m $, $\hbar$ is the Planck's constant, $a_s$ is the $s$-wave scattering length, and $j_1$ is the spherical Bessel function. 
The amplitude is chosen to be such that with vanishing of the range $r_0$ of the interatomic potential, $\lim_{r_0\to0}\nu_{\bf p}=g$, which corresponds to the local-interaction limit, see also \ref{app:A} for details.

\begin{figure}[t]
  \includegraphics[width=\columnwidth]{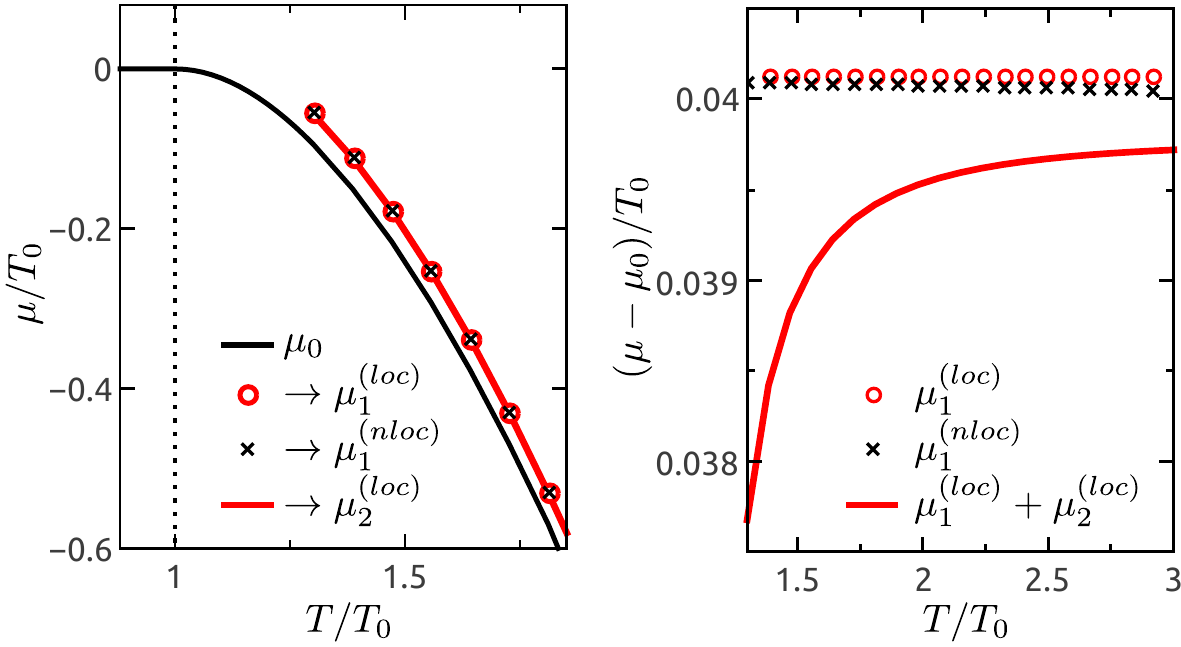}    
  \caption{\label{fig:1}
    Temperature dependencies of the chemical potentials (the arrows point to the highest limiting orders employed in the calculation) and the corresponding corrections obtained within TPT for a weakly interacting Bose gas.
    The parameters are $n=10^{12}$~cm$^{-3}$, $a_s=100a_{\rm B}$, and $r_0=\Lambda_0/20\approx 1300 a_{\rm B}$.
    }
\end{figure}

In Fig.~\ref{fig:1} one can see the temperature dependencies of the chemical potential up to quadratic corrections (both for local and nonlocal potentials) and with account for non-locality of interaction.
Here and below, the evaluation of the results in the numerical procedure is limited from below in $T$ by the condition $|{\cal O}_2(T^*)| = 0.1|{\cal O}_1(T^*)|$, where ${\cal O}$ is the thermodynamic quantity of interest (more specifically, for consistency of thermodynamic relations, we restrict calculations of the chemical potential~$\mu$ and other thermodynamic quantities by the condition for the pressure~$P$, see below). In this way, the characteristic temperature $T^*$ can be associated with the region of the breakdown of the TPT applicability. For the given $n$ and $a_s$, which are chosen close to the typical values in dilute gases of alkali-metal atoms and shown in Fig.~\ref{fig:1}, $T^*\approx1.3T_0$.

In accordance with Eqs.~\eqref{eq:mu}--\eqref{eq:muloc} and Ref.~\cite{chempot2017ltp} for $\mu_0(T)$ at fixed density~$n$, we note that the linear TPT correction yields a monotonous shift (an increase) of the chemical potential in comparison with the non-interacting limit.
The quadratic correction results in a slight decrease of the chemical potential $\mu$ toward $\mu_0$. 
One can also see that the nonlocal interaction suppresses the value of the linear correction, as can also be noticed from the analysis of Eq.~\eqref{eq:N1}.
However, the effect becomes noticeable only in the case of long-range interatomic potentials with $r_0\gtrsim 10^3 a_{\rm B}$ (e.g., in case of ultracold Rydberg atoms, where $a_{\rm B}$ is the Bohr radius). Otherwise, for typical cases of dilute gases and the potentials with $r_0\ll 10^3a_{\rm B}$ (e.g., ultracold alkali metal atoms in the ground state), these corrections are negligible in the studied range of the TPT applicability (i.e., at $T\gtrsim T^*$). Therefore, the interaction effects above the transition temperature can be described well by the local approximation in case of ultracold gases of atoms with moderate range of interatomic potentials in real space (see also \ref{app:A} for more details).

\subsection{Pressure}
Next, let us analyze the pressure $P$,
\begin{equation}\label{eq:defP}
	P =	- {\Omega}/{V}.
\end{equation}
Before employing Eqs.~\eqref{eq:Om}--\eqref{eq:Om2} to obtain $P$ from Eq.~\eqref{eq:defP},
it is necessary to additionally expand $\Omega$ into series due to the fact that the chemical potential~$\mu$ 
contains the functional dependence on the interaction as well. Thus, we find that
\begin{eqnarray}
 	P(\mu) 
 	= P_0(\mu_0) + P_1(\mu_0)
 	+ P_2(\mu_0) + \dots
 	\\
 	 P_0(\mu_0) = - { \Omega_0(\mu_0) }/{ V },
 	\\
     P_1(\mu_0) = n \mu_1 - \frac1{ 2V^2 } \sum\limits_{\bf{pp}^\prime} \left[ \nu_0 + \nu_{ \bf{p} \bf{p}^\prime } \right] f_p f_{p^{\prime}} ,
 	\\
 	 P_2(\mu_0) = \left( \frac{ \partial N_0 ( \mu_0 ) }{ \partial \mu_0 } \right) _{T,V} 
 	 \frac{	\mu_1^2 }{ 2V }
 	 + N_1 ( \mu_0 ) \frac{	\mu_1 }{ V }
 	 \nonumber
 	 \\
 	 \qquad\quad
 	 + P_2^\prime(\mu_0) + P_2^{\prime\prime}(\mu_0) + n	\mu_2,
\end{eqnarray}
where
\begin{eqnarray*}
    P_2^\prime(\mu_0) 
    = \frac{ \beta }{ 2V^3 } 
    \sum\limits_{{\bf p}_1{\bf p}_2{\bf p}_3} 
    \left( 
        \nu_0^2 
        + 2 \nu_0 \nu_{{\bf p}_2{\bf p}_3} 
        + \nu_{{\bf p}_1{\bf p}_3} \nu_{{\bf p}_2{\bf p}_3} 
    \right)
    \\
    \qquad\qquad\times
    f_{p_1} f_{p_2} f_{p_3} \left( f_{p_3} + 1 \right) ,
    \\
    P_2^{\prime\prime}(\mu_0) 
    = \frac{ \beta }{ 4V^3 } 
    \sum\limits_{ {\bf p}_1 \dots {\bf p}_4 } 
    \left( 
        \nu_{{\bf p}_1{\bf p}_3}^2 
        + \nu_{{\bf p}_1 {\bf p}_3} 
        \nu_{{\bf p}_1{\bf p}_4} 
    \right) 
    \\
    \qquad\qquad\times
    f_{p_1} f_{p_2} \left( f_{p_3} + 1 \right) \left( f_{p_4} + 1 \right) 
    \delta_ { {\bf p}_1 + {\bf p}_2 , 
    {\bf p}_3 + {\bf p}_4 }.
\end{eqnarray*}

In case of local interaction, the gas pressure can be written as follows:
\begin{equation}\label{eq:Ploc}
	P_{1}^{(loc)}
	=
	n^2
	\nu_0
,~
	P_{2}^{(loc)}(\mu_0)
	=
	P_{2\,loc}^{\prime\prime}(\mu_0)
	+n\mu_{2}^{(loc)}.
\end{equation}

\begin{figure}[t]
  \includegraphics[width=\columnwidth]{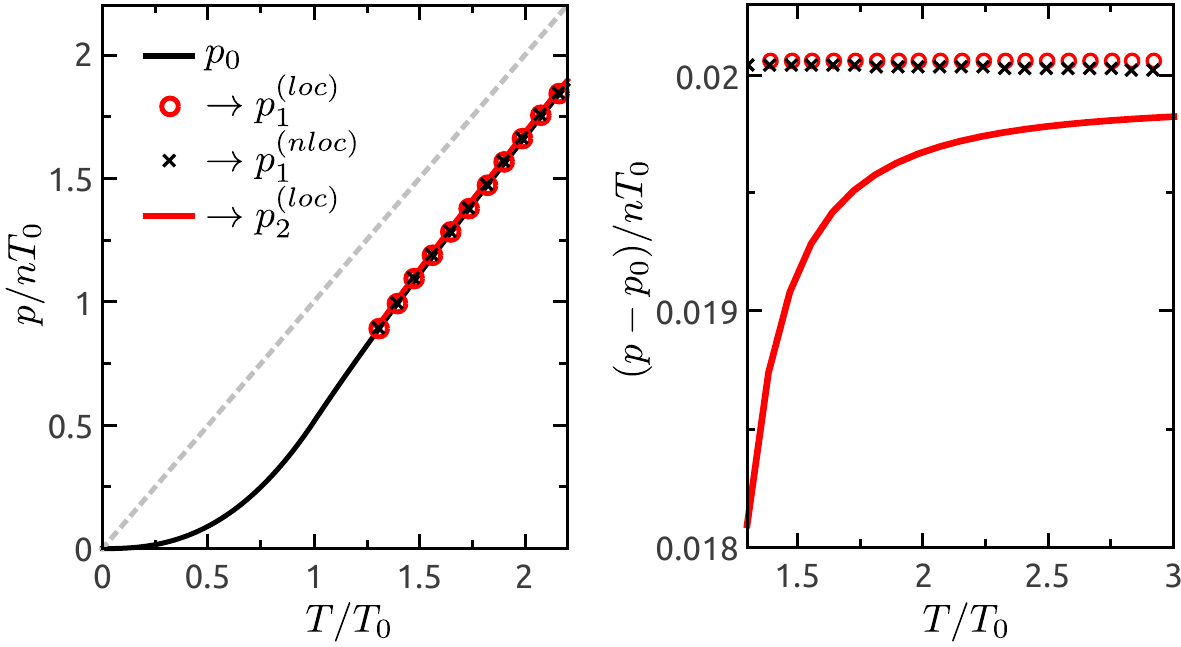}    
  \caption{\label{fig:2}
    Temperature dependencies of the pressure and the corresponding corrections obtained within TPT for a weakly interacting Bose gas.
    The dashed line corresponds to the classical ideal gas.
    The parameters are $n=10^{12}$~cm$^{-3}$, $a_s=100a_{\rm B}$, and $r_0=\Lambda_0/20$.
    }
\end{figure}

The numerical results for the pressure are shown in Fig.~\ref{fig:2}. 
These have similar qualitative peculiarities to those that are shown for the chemical potential: the linear TPT correction is temperature independent and positive; the quadratic one is negative but small; the nonlocality of the interaction has a minor impact on the corrections for the typical values of $r_0$ and $n$ above $T^*$.

\subsection{Entropy and specific heat}
To determine the entropy, we employ the Maxwell relation
\begin{equation}\label{eq:defS}
    S = -
    \left(
    \frac{\partial\Omega}{\partial T}
    \right)_{\mu,V}.
\end{equation}
By plugging the explicit expressions~\eqref{eq:Om}--\eqref{eq:Om2} for $\Omega$ into Eq.~\eqref{eq:defS} and additionally expanding dependencies for the chemical potential~$\mu$ into series in $\beta\nu$, we obtain
\begin{eqnarray}
    S(\mu) 
	 = S_0(\mu_0) + S_1(\mu_0) + S_2(\mu_0) + \dots,
	\label{eq:S}
	\\ 
	S_0(\mu_0) 
	 = \sum\limits_{\bf p} 
	 \left[ 
	  \left( 1 + f_p \right) \ln{ \left( 1 + f_p \right) } - f_p \ln{ f_p } 
	 \right],
	\\
	S_1(\mu_0) 
	= \left( 
	 \frac{ \partial N_0 ( \mu_0 ) }{ \partial T } 
	 \right)_{\mu_0,V} \mu_1
	 +S_{1}''(\mu_0)
	 ,\label{eq:S1}
\end{eqnarray}
where $S_{0}(\mu_{0})$ is the entropy of an ideal Bose gas (combinatorial definition) and 
\begin{eqnarray*}
	S_{1}''(\mu_0)
	=
	 - \frac{ \beta^2 }{ V } \sum\limits_{{\bf pp}^\prime} 
	\left[ \nu_0 + \nu_{{\bf p} {\bf p}^\prime } \right] 
	\left( \varepsilon_p - \mu \right) f_{p^{\prime}} f_p 
	\left( f_p + 1 \right).
\end{eqnarray*}
The quadratic correction can be written as
\begin{eqnarray}
    S_2(\mu_0)
	= \left( \frac{ \partial^2 S_0 (\mu_0) }{ \partial \mu_0^2 } \right)_{T,V} 
	\frac{	\mu_1^2 }2 + 
	\left( \frac{ \partial S_{1}'' (\mu_0) }{ \partial \mu_0 } 
	\right)_{T,V} \mu_1
	\nonumber
\\
    \qquad~
	+ S_2^{\prime}(\mu_0)
	+ S_2^{\prime\prime}(\mu_0) +	\left( \frac{ \partial N_0 (\mu_0) }{ \partial T } \right) _{\mu_0,V} \mu_2,\label{eq:S2}
\end{eqnarray}
where
\begin{eqnarray*}
    S_2^{\prime}(\mu_0)
	= - \beta \Omega_2^{\prime}(\mu_0)
	+ \frac{ \beta^3 }{ 2V^2 } \sum\limits_{{\bf p}_1{\bf p}_2{\bf p}_3} 
	\left[ 
	 \nu_0^2  + 2 \nu_0 \nu_{{\bf p}_2{\bf p}_3}
	\right.
\\
    ~~
    \left.
     + \nu_{{\bf p}_1{\bf p}_3} \nu_{{\bf p}_2{\bf p}_3} 
    \right] 
  f_{p_1} f_{p_2} f_{p_3} ( f_{p_3} + 1 )
  \left[ \left( f_{p_1} + 1 \right) \left( \varepsilon_{p_1} - \mu_0 \right)
  \right.
\\ 
    ~~
   \left.
  + \left( f_{p_2} + 1 \right) \left( \varepsilon_{p_2} - \mu_0 \right) 
  + \left( 2 f_{p_3} + 1 \right) \left( \varepsilon_{p_3} - \mu_0 \right) \right],
\\
 	S_2^{\prime\prime}(\mu_0)
	= - \beta \Omega_2^{\prime\prime}(\mu_0)
\\
  +
	\frac{ \beta^3 }{ 4V^2 } 
	\sum\limits_{ {\bf p}_1 \dots {\bf p}_4 } 
	\left[ \nu_{{\bf p}_1{\bf p}_3}^2 + \nu_{{\bf p}_1{\bf p}_3} \nu_{{\bf p}_1{\bf p}_4} 
	\right] 
	f_{p_1} f_{p_2} \left( f_{p_3} + 1 \right)
\\
    \qquad\times
	\left( f_{p_4} + 1 \right) 
	\delta_{{\bf p}_1 + {\bf p}_2 , {\bf p}_3 + {\bf p}_4 }
	\\
	\qquad\times  
	\left[ \left( f_{p_1} + 1 \right) 
	\left( \varepsilon_{p_1} - \mu_0 \right) 
	+ \left( f_{p_2} + 1 \right) 
	\left( \varepsilon_{p_2} - \mu_0 \right) 
	\right.
	\\
	\left.
	\qquad~~  + f_{p_3}
	\left( \varepsilon_{p_3} - \mu_0 \right) + f_{p_4} 
	\left( \varepsilon_{p_4 } - \mu_0 \right) \right].
\end{eqnarray*}

\begin{figure}[t]
  \includegraphics[width=\columnwidth]{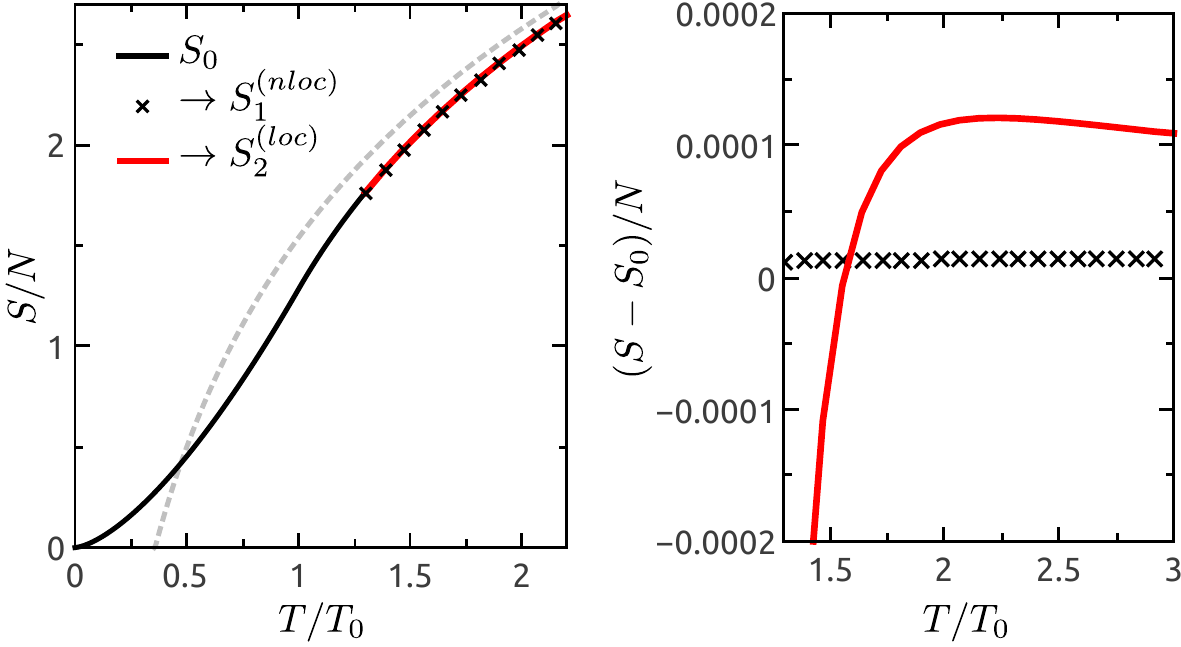}    
  \caption{\label{fig:3}
    Temperature dependencies of the entropy and the corresponding corrections obtained within TPT for a weakly interacting Bose gas.
    The dashed line corresponds to the classical ideal gas.
    The parameters are $n=10^{12}$~cm$^{-3}$, $a_s=100a_{\rm B}$, and $r_0=\Lambda_0/20$.
    }
\end{figure}

Note that in the limit of local interaction, the first-order
TPT correction to the entropy \eqref{eq:S1} vanishes,
\begin{equation}\label{eq:S_loc1}
	S_{1}^{(loc)}(\mu_0) = 0,
\end{equation}
and the second-order term \eqref{eq:S2} can be written in a more compact form as follows:
\begin{equation}\label{eq:S_loc}
	S_{2}^{(loc)} (\mu_0)
	=
	S_{2}''^{(loc)}(\mu_0) + 
	\left( \frac{ \partial N_0 (\mu_0) }{ \partial T } 
	\right)_{\mu_0,V} \mu_{2}^{(loc)}.
\end{equation}
The results for both local and nonlocal interaction parameterizations are summarized in Fig.~\ref{fig:3}. 

Now, let us determine the specific heat capacity at constant volume as
\cite{Landau5eng}
\begin{equation}\label{eq:def_Cv}
	C_V
	= T
	\left[ \left( \frac{ \partial S }{ \partial T } \right)_{\mu,V} 
	- 
	\left( \frac{ \partial N }{ \partial T } \right)_{\mu,V} ^2
	\left( \frac{ \partial N }{ \partial \mu } \right) ^{-1}_{T,V} 
	\right].
\end{equation}
Since the derivatives enter Eq.~\eqref{eq:def_Cv} in the nonlinear manner, the expression is necessary to additionally expand into the TPT series.
Hence, it is convenient to express the specific heat in the same way as it is done for the chemical potential, 
\begin{eqnarray*}
 C_V(\mu)
	=
	C_{V0}(\mu_0) + C_{V1}(\mu_0) + C_{V2}(\mu_0) + \dots,
 \\
 C_{V1}(\mu_0)
	\equiv
	\sum\limits_{\bf p} 
	\left. 
	\left( \frac{ \partial C_V(\mu) }{ \partial \nu_{\bf p}} 
	\right)_{T,V} 
	\right|_{0} 
	\nu_{\bf p},
 \\
 C_{V2}(\mu_0)
	\equiv
	\frac12 \sum\limits_{{\bf pp}^\prime} 
	\left. \left( \frac{ \partial^2 C_V(\mu) }{ \partial \nu_{\bf p} 
	\partial \nu_{{\bf p}^\prime} } \right)_{T,V} \right|_{0} 
	\nu_{\bf p} \nu_{{\bf p}^\prime}.
\end{eqnarray*}

For the sake of clarity of the results given in this subsection, we do not provide explicit expressions for $C_{V1}$ and $C_{V2}$ in general case, which can be found in \ref{app:B}, but specify their rather compact form in case of local interaction. Similarly to the entropy, the linear term vanishes,
\begin{equation}\label{eq:CV_loc1}
	C_{V1}^{(loc)}(\mu_0) = 0,
\end{equation}
and the quadratic TPT correction can be written as
\begin{equation}\label{eq:CV_loc}
	C_{V2}^{(loc)} (\mu_0)
	=
	C_{V2}''^{(loc)}(\mu_0) 
	+ \left( 
	\frac{ \partial C_{V0} (\mu_0) }{ \partial T } 
	\right)_{\mu_0,V} \mu_{2}^{(loc)},
\end{equation}
where the explicit form for $C_{V2}''(\mu_0)$ can be found in \ref{app:B}.

\begin{figure}[t]
  \includegraphics[width=\columnwidth]{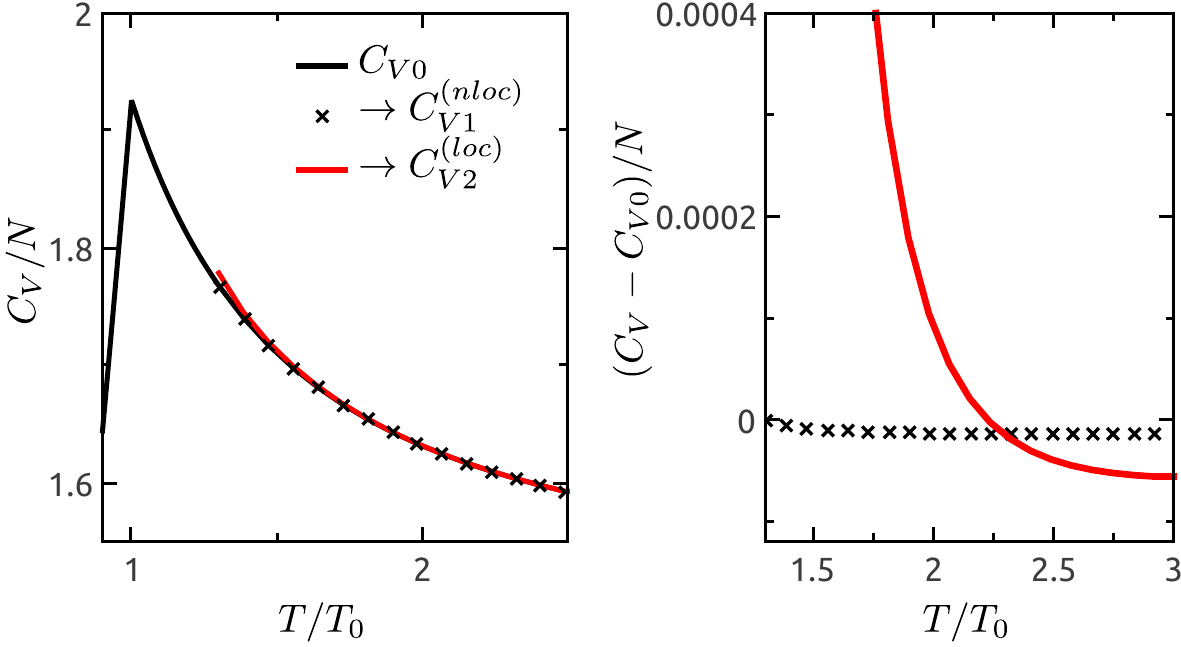}    
  \caption{\label{fig:4}
    Temperature dependencies of the specific heat and the corresponding corrections obtained within TPT for a weakly interacting Bose gas.
    The parameters are $n=10^{12}$~cm$^{-3}$, $a_s=100a_{\rm B}$, and $r_0=\Lambda_0/20$.
    }
\end{figure}

From Fig.~\ref{fig:4} we see that the TPT corrections to the specific heat are relatively small in the range of applicability of the theoretical approach and for the gas parameters typical for the experiments with ultracold atoms.

\section{Conclusion}
We developed and applied the thermodynamic perturbation theory to the weakly interacting Bose gases above the transition temperature to the BEC state. The temperature dependencies of main thermodynamic quantities were analyzed both analytically and numerically for contact (local) and non-local interaction potentials. It was demonstrated that the local interaction specified by the scattering length describes fairy well the interacting Bose gas above the transition temperature.   
At the same time, the effects of nonlocality give noticeable contributions to thermodynamics of weakly interacting gases as soon as the characteristic range of atomic potentials approaches the mean interparticle distance.
This is important in view of studies of cold gases of atoms with long-range interactions, e.g., prepared in the Rydberg states.

The results also show that the interaction gives the largest contribution to the chemical potential $\mu$ and the pressure $P$.
For the local interaction, which is determined by a single quantity --- the scattering length $a_s$ --- the linear corrections to the entropy $S$ and the specific heat $C_V$ vanish.
It is shown that the entropy $S$ and the specific heat $C_V$ are described well within the ideal-gas approximation for the typical experimental parameters of trapped atomic gases.
The obtained dependencies (in particular, for the chemical potential) also provide a reliable asymptotic behavior above the transition temperature for existing 
\ao{theoretical
(mean-field and quantum-Monte-Carlo) 
as well as experimental}
studies, see, e.g., Ref.~\cite{Stringari2020PRL,Ota2020pre} and references therein.

\section*{Acknowledgements}
The authors acknowledge funding by the National Research Foundation of Ukraine, 
Grant No.~0120U104963.
and the Ministry of Education and Science of Ukraine, Research Grant No.~0120U102252.

\begin{appendix}

\section{Non-local interaction: structure in the momentum space}\label{app:A}
 
Here, we discuss the observed smallness of the nonlocal effects in the TPT corrections to the main thermodynamic quantities under study.
\begin{figure}[t]
  \includegraphics[width=\columnwidth]{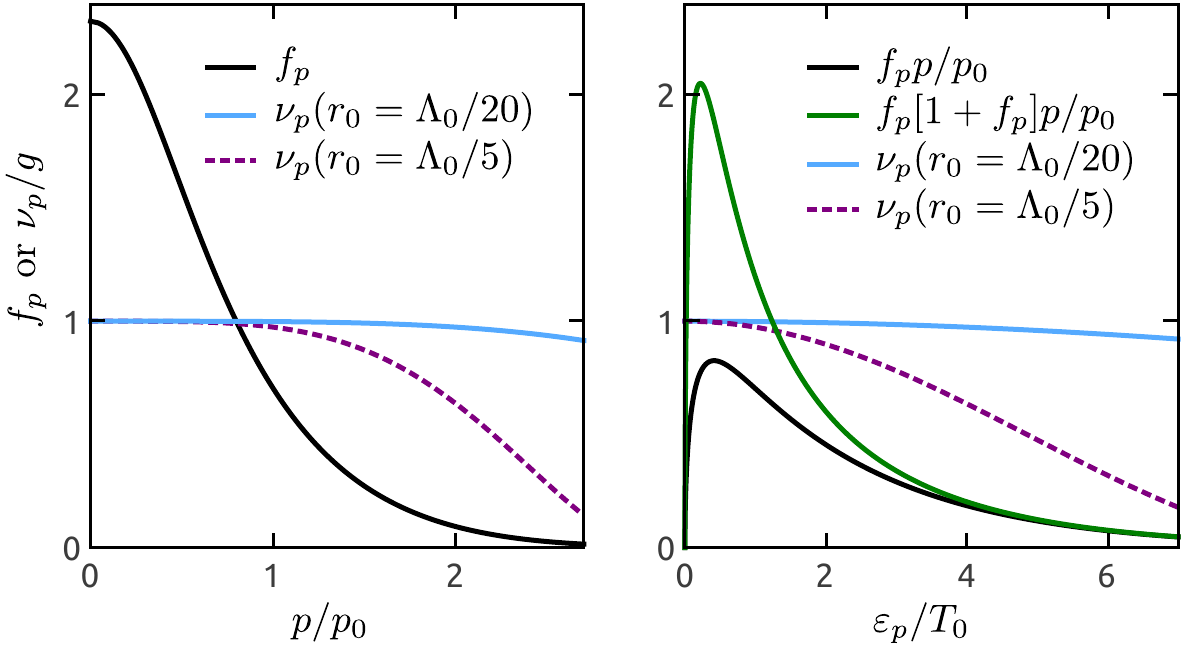}    
  \caption{\label{fig:5}
    Profiles of the interaction potentials together with the structure of the distribution function of a gas at $T=1.9T_0$.
    }
\end{figure}
From Fig.~\ref{fig:5} one can see that for the central case of the study, where the relatively small range~$r_0=\Lambda_0/20$ of the potential~$\nu_p$, its profile is almost flat compared to the change of the Bose-Einstein distribution function and related quantities at characteristic temperatures of the gas. This means that the local approximation remains accurate in the framework of the TPT. 
This contrasts to the studies of a Bose condensed gas at zero temperature within the completely consistent quadratic approximation \cite{Tolmachev,Bulakhov2018}, where the main equation ensuring the minimum of the thermodynamic potential has no solution even after applying the well-known renormalization procedure for the coupling constant given by the scattering length \cite{Pethick, Stringari}.
For the nonlocal interaction potential, this problem can be accurately solved, while the single-particle excitation spectrum acquires a gap \cite{Bulakhov2018}.
 
At the same time, from Fig.~\ref{fig:5} it becomes clear that, as soon as the interaction becomes long-ranged, i.e., the characteristic range~$r_0$ is of the same order as the mean interparticle distance, the nonlocal effects become more pronounced. Presumably, these effects can be probed with the ultracold Rydberg gases.

\section{TPT corrections for the specific heat: explicit form}\label{app:B}
In general case, the linear TPT correction to the specific heat can be written in the following form:
\begin{eqnarray*}
    C_{V1}(\mu_0)
	 =
	\left.
	 (\partial_\mu C_{V0})
	\right|_{0}
	\mu_1
	+C_{V1}''(\mu_0),
\end{eqnarray*}
where
\begin{eqnarray*}
	C_{V1}''(\mu_0)
	&=&	T\left[ 
	 \partial_T S_{1}''
	- 2 (\partial_T N_1)
	(\partial_T N_0)
	(\partial_\mu N_0)^{-1}
		\right.
	\\
	&&\left.\left.
		+(\partial_T N_0)^2
		(\partial_\mu N_1)
		(\partial_\mu N_0)^{-2}
	\right]\right|_0
\end{eqnarray*}
and we introduced the compact notations $\partial_T A \equiv (\partial A/\partial T)_{\mu,V}$ and $\partial_\mu A \equiv (\partial A/\partial \mu)_{T,V}$.

The quadratic TPT correction can be written as
\begin{eqnarray*}
	C_{V2}(\mu_0)
	&=&
	\left.
	(\partial^2_\mu C_{V0})
	\right|_{0}
	\frac{\mu_1^2 }2
	+ C_{V2}^{\prime}(\mu_0)
	+ C_{V2}^{\prime\prime}(\mu_0)
	\\
	&&+
	\left.
	(\partial_\mu C_{V1\nu})
	\right|_{0}
	\mu_1
	+
	\left.
	 (\partial_\mu C_{V0})
	\right|_{0}
	\mu_2,
\end{eqnarray*}
where
\begin{eqnarray*}
	C_{V2}^{\prime}(\mu_0)
	=
	T\left[
	 \partial_T S_{2}^{\prime}
	 -2
	 (\partial_T N_{2}^{\prime})
	 (\partial_T N_{0})
	 (\partial_\mu N_{0})^{-1}
	\right.
	\\
	\qquad
	+ 2 
	 (\partial_T N_{1})
	 (\partial_T N_{0})
	 (\partial_\mu N_{1})
	 (\partial_\mu N_{0})^{-1}
	\\
	\qquad
	-
	(\partial_T N_{1})^2
	(\partial_\mu N_{0})^{-1}
	- 
	(\partial_T N_{0})^2
	(\partial_\mu N_{1})^2
	(\partial_\mu N_{0})^{-3}
	\\
	\qquad
	+
	\left.\left.
	 (\partial_T N_{0})^2
	 (\partial_\mu N_{2}')
	 (\partial_\mu N_{0})^{-2}
	\right]
	\right|_0
\end{eqnarray*}
and
\begin{eqnarray*}
	C_{V2}^{\prime\prime}(\mu_0)
	&=&
	T\left[
	 \partial_T S_{2}^{\prime\prime}
	 -2
	 (\partial_T N_{2}^{\prime\prime})
	 (\partial_T N_{0})
	 (\partial_\mu N_{0})^{-1}
	\right.
	\\
	&&+
	\left.\left.
	 (\partial_T N_{0})^2
	 (\partial_\mu N_{2}'')
	 (\partial_\mu N_{0})^{-2}
	\right]
	\right|_0.
\end{eqnarray*}

\end{appendix}

\section*{References}
\bibliography{tpt_bose_gas} 
\end{document}